\documentclass[11pt,a4paper]{article}
\usepackage{jheppub}

\usepackage{graphicx}
\usepackage{amsmath}
\usepackage{epsfig}
\usepackage{hhline}
\usepackage{soul}
\usepackage{tabularx,pifont,multirow}
\usepackage{enumitem}
\usepackage{colordvi}

\RequirePackage{xspace}

\input babarsym

\newcommand\trule{\rule{0pt}{3.0ex}}

\def\K       {\ensuremath{K}\xspace}

\def\Kminus {\ensuremath{ K_{-} }\xspace}
\def\Kplus {\ensuremath{ K_{+} }\xspace}

\def\Bminustilde {\ensuremath{ \tilde{B}_{-} }\xspace}
\def\Bplustilde {\ensuremath{ \tilde{B}_{+} }\xspace}
\def\Bminus {\ensuremath{ B_{-} }\xspace}
\def\Bplus {\ensuremath{ B_{+} }\xspace}

\def\jpsi {\ensuremath{ J/\psi }\xspace}

\def\dm {\ensuremath{ \Delta m}\xspace}
\def\dt {\ensuremath{ \Delta t }\xspace}
\def\dtplus {\ensuremath{ \Delta t_+ }\xspace}
\def\dtau {\ensuremath{ \Delta \tau }\xspace}
\def\dtrec {\ensuremath{ \Delta t_{\rm rec} }\xspace}

\def\SpLpKs  {\ensuremath{ S^+_{\ellp,\KS} }\xspace}
\def\SpmLpKs  {\ensuremath{ S^\pm_{\ellp,\KS} }\xspace}
\def\SpLmKs {\ensuremath{ S^+_{\ellm,\KS} }\xspace}
\def\SmLpKs  {\ensuremath{ S^-_{\ellp,\KS} }\xspace}
\def\SmLmKs {\ensuremath{ S^-_{\ellm,\KS} }\xspace}
\def\SpLpKl  {\ensuremath{ S^+_{\ellp,\KL} }\xspace}
\def\SpLmKl {\ensuremath{ S^+_{\ellm,\KL} }\xspace}
\def\SmLpKl  {\ensuremath{ S^-_{\ellp,\KL} }\xspace}
\def\SmLmKl {\ensuremath{ S^-_{\ellm,\KL} }\xspace}
\def\CpLpKs  {\ensuremath{ C^+_{\ellp,\KS} }\xspace}

\def\CpLmKs {\ensuremath{ C^+_{\ellm,\KS} }\xspace}
\def\CmLpKs  {\ensuremath{ C^-_{\ellp,\KS} }\xspace}
\def\CmLmKs {\ensuremath{ C^-_{\ellm,\KS} }\xspace}
\def\CpLpKl  {\ensuremath{ C^+_{\ellp,\KL} }\xspace}
\def\CpLmKl {\ensuremath{ C^+_{\ellm,\KL} }\xspace}
\def\CmLpKl  {\ensuremath{ C^-_{\ellp,\KL} }\xspace}
\def\CmLmKl {\ensuremath{ C^-_{\ellm,\KL} }\xspace}
\def\SpBzKs  {\SpLpKs}

\def\SmBzKs  {\SmLpKs}

\def\CpBzKs  {\CpLpKs}

\def\CmBzKs  {\CmLpKs}

\def\DeltaSpmT {\ensuremath{ \Delta S^\pm_{\rm  T} }\xspace}
\def\DeltaCpmT {\ensuremath{ \Delta C^\pm_{\rm  T} }\xspace}

\def\DeltaSpmCP {\ensuremath{ \Delta S^\pm_{\rm  CP} }\xspace}
\def\DeltaCpmCP {\ensuremath{ \Delta C^\pm_{\rm  CP} }\xspace}

\def\DeltaSpmCPT {\ensuremath{ \Delta S^\pm_{\rm  CPT} }\xspace}
\def\DeltaCpmCPT {\ensuremath{ \Delta C^\pm_{\rm  CPT} }\xspace}

\def\DeltaSmT {\ensuremath{ \Delta S^-_{\rm  T} }\xspace}
\def\DeltaCmT {\ensuremath{ \Delta C^-_{\rm  T} }\xspace}

\def\DeltaSmCP {\ensuremath{ \Delta S^-_{\rm  CP} }\xspace}
\def\DeltaCmCP {\ensuremath{ \Delta C^-_{\rm  CP} }\xspace}

\def\DeltaSmCPT {\ensuremath{ \Delta S^-_{\rm  CPT} }\xspace}
\def\DeltaCmCPT {\ensuremath{ \Delta C^-_{\rm  CPT} }\xspace}

\def\DeltaSpT {\ensuremath{ \Delta S^+_{\rm  T} }\xspace}
\def\DeltaCpT {\ensuremath{ \Delta C^+_{\rm  T} }\xspace}

\def\DeltaSpCP {\ensuremath{ \Delta S^+_{\rm  CP} }\xspace}
\def\DeltaCpCP {\ensuremath{ \Delta C^+_{\rm  CP} }\xspace}

\def\DeltaSpCPT {\ensuremath{ \Delta S^+_{\rm  CPT} }\xspace}
\def\DeltaCpCPT {\ensuremath{ \Delta C^+_{\rm  CPT} }\xspace}

\newcommand{\phm}{\ensuremath{\phantom{-}}}

\newcommand{\ket}[1]{\ensuremath{|{#1}\rangle}}
  
\def\beq{\begin{equation}}
\def\eeq{\end{equation}}
\def\bea{\begin{eqnarray}}
\def\eea{\end{eqnarray}}
\def\bq{\begin{quote}}
\def\eq{\end{quote}}
\def\ben{\begin{enumerate}}
\def\een{\end{enumerate}}
\def\nn{\nonumber}

\def\twoDLL  {\ensuremath{-2\Delta\ln{\cal L}}\xspace}
\def\CL {\ensuremath{ \rm C.L. }\xspace}

\long\def\inst#1{\par\nobreak\kern 4pt\nobreak
  {\it #1}\par\vskip 10pt plus 3pt minus 3pt}

\title{\large \bf
\boldmath
\Large
Time Reversal Violation 
from the entangled $\Bz\Bzb$ system
} 

\author[a,b]{J.~Bernab\'eu}
\affiliation[a]{Department of Theoretical Physics, Universitat de Val\`encia, E-46100 Burjassot, Spain}
\affiliation[b]{IFIC, Universitat de Val\`encia-CSIC, E-46071 Val\`encia, Spain}
\author[b]{F.~Mart\'inez-Vidal}
\author[b]{P.~Villanueva-P\'erez}

\abstract{

\noindent
We discuss the concepts and methodology to implement an experiment
probing directly Time Reversal (\T) non-invariance, without any
experimental connection to \CP violation, by the exchange of {\it in} and
{\it out} states. The idea relies on the $\Bz\Bzb$ entanglement and decay
time information available at B factories. The flavor or \CP tag of the
state of the still living neutral meson by the first decay of its
orthogonal partner overcomes the \mbox{problem} of irreversibility for
unstable systems, which prevents direct tests of \T with incoherent particle states.
\T violation in the time evolution between the two decays
means experimentally a difference between
the intensities for the time-ordered ($\ellp X$, $\jpsi\KS$) and ($\jpsi\KL$, $\ellm X$) decays, and three other
independent asymmetries. 
The proposed strategy has been \mbox{applied}
to simulated data samples of similar size and features to those
currently available, from which we estimate the significance of the
expected discovery to reach many standard \mbox{deviations}.

}

\begin{document}

\date{\today}

\maketitle

\newpage


\setcounter{footnote}{0}


\section{ \boldmath Introduction}
\label{sec:intro}
Violation of the \CP symmetry 
has been observed in \K and \B meson decays~\cite{ref:Christenson,ref:BannerAdlerApostolakis,ref:mixingInducedCP-Bs,ref:directCP-Bs}. 
In the Standard Model (SM) the mechanism of \CP violation in weak interactions arises from the joint effect of the three non-vanishing mixing angles and the single irreducible phase in the 
3-family Cabibbo-Kobayashi-Maskawa (CKM) mixing matrix. 
The existence of this matrix conveys the fact that the quarks that participate in weak processes are a linear combination of mass eigenstates.
This mechanism has been confirmed to be the dominant by experiments probing \CP violation, 
particularly with studies involving \B decays~\cite{ref:HeavyQuarksRept}.  

In the context of local quantum field theories with Lorentz invariance and Hermiticity, the \CPT theorem~\cite{ref:Ludgers} ensures a theoretical 
constraint between the \CP and \T symmetries. 
Although all present tests of \CPT invariance confirm the validity of this symmetry,
particularly in the neutral-kaon system where there are strong limits~\cite{ref:CPTtests,ref:pdg2010},
the theoretical connection between \CP and \T does not imply an experimental
identity, except for processes which are identical under \CPT transformation.
Moreover, it is 
worthwhile
to search for direct evidence of \T non-invariance, 
with neither experimental nor theoretical connection to \CP violation and \CPT invariance.
There is at present no existing result that clearly shows \T violation in this sense~\cite{ref:Wolfenstein}.

In the case of transition processes, 
due to the antiunitarity of the operator implementing the symmetry, 
\T invariance requires that the likelihood for reaction {\it in} $\to$ {\it out} equals that of {\it out} $\to$ {\it in} once
the initial configurations, namely {\it in} in one case and {\it out} in the other, have been precisely prepared. 
However, the likelihood of the time reversed version of the process to happen is very low or impractical in some cases, 
like for unstable systems. 
This explains why \T is much more difficult to study directly than \P, \C, and \CP.

Since the SM is \CPT invariant, it predicts 
\T-violating effects in parallel to 
each \CP-violation effect that arises due to the interference of amplitudes with different weak phases.
These may appear in three different ways: \T violation in decay;
\T violation in the mixing of neutral states; and \T violation that arises from the interference
between decay with and without mixing.

\T violation matched to \CP violation in decay has not been observed, because the difficulties of the preparation 
of the time reversed decay process.
Let us see the example~\cite{ref:QuinnDiscrete} of the rare weak decay of neutral-\B mesons 
to $\Kp\pim$,
for which direct \CP violation is well established~\cite{ref:directCP-Bs} (different decay rates $R_1$ and $R_2$ for $\Bz\to\Kp\pim$ and $\Bzb\to\Km\pip$, 
respectively) due to the ability of B factories to produce hundreds of millions of \B mesons. \T violation as implied by this result, 
combined with \CPT invariance, tells us that the rates for the inverse processes $\Kp\pim\to\Bz$ and $\Km\pip\to\Bzb$ should be $R_2$ and $R_1$, 
respectively. However, there is little chance to measure these inverse rates and check directly this prediction since
the weak interaction production mechanism is highly suppressed ($\Bz\to\Kp\pim$ branching ratio of order $10^{-5}$) and
the strong interaction would completely swamp the feeble weak process.

\T violation associated to \CP violation in the mixing has been experimentally analyzed in \K~\cite{ref:Angelopoulos} and \B mesons~\cite{ref:TviolationBs}.
Here one looks whether the rate for a neutral-\K (\B) meson tagged at its production as \Kz (\Bz) and identified afterwards as \Kzb (\Bzb)
is equal to the rate for the neutral particle tagged at its production as \Kzb (\Bzb) and identified later as \Kz (\Bz).
Any difference in this case is both \CP and \T violating, because \CP and \T are experimentally identical for this process.
The experimental results for kaons yielded a \T-violating difference in these rates. 
Such a difference is proportional not only to the \T-violating term of the $\Kz\Kzb$ matrix
that defines the mass eigenstates in terms of the flavor eigenstates, 
but also to the width difference $\Delta \Gamma$ between the two mass eigenstates,
thus \T non-invariance would not be present in the limit $\Delta \Gamma \to 0$. 
Therefore, this asymmetry shows \T violation proportional to $\Delta \Gamma$, time independent,
experimentally identical to \CP violation, thus it is not an independent \T non-invariance test as one might like~\cite{ref:Wolfenstein}. 
In the $\Bz\Bzb$ system no asymmetry has been yet found, as expected within the SM since in this 
case $\Delta \Gamma$ almost vanishes~\cite{ref:TviolationBs-theo}.

The largest 
\CP-violating asymmetry in Nature has been found between the 
rate for $\Bz\to\jpsi\KS$ (and other similar \CP-odd $\ccbar\KS$ final states, e.g.,
$\psi(2S)\KS$, $\chic1\KS$, or the \CP-even $\jpsi\KL$ final state) 
and the \CP-conjugate rate for \Bzb to decay to the same \CP-eigenstate, which is originated in the interference between the 
time-dependent decay amplitudes with and 
without mixing~\cite{ref:mixingInducedCP-Bs,ref:HeavyQuarksRept}. 
Here, \CP violation arises because the mixing phase, 
i.e., the relative phase between the complex parameters defining 
the mass eigenstates in terms of the flavor eigenstates~\cite{ref:sin2beta}
minus the relative phase of the ratio of the amplitude for the decay and its \CP conjugate, 
does not vanish. In the SM this phase
difference is $2\beta$, where $\beta$ is the angle between the $V_{\c\d}V_{\c\b}^*$ and $V_{\t\d}V_{\t\b}^*$ sides of the CKM unitarity triangle.
At B factories the experimental studies are performed using the two \B mesons produced in the antisymmetric coherent state from the \FourS decay.
One neutral-\B meson decays into a definite flavor state and the other is reconstructed in the \CP-eigenstate final state of interest
with a given decay time difference $\dt=t_{\CP}-t_{\rm flavor}$.
The measured asymmetry is large, 
proportional to $\sin2\beta \approx 0.7$, 
a time-odd dependent function
that reverses sign between \Bz and \Bzb tagged events, and between $\Bz\to\jpsi\KS$
and $\Bz\to\jpsi\KL$ events 
with the same flavor tag. This is the observed \CP-violating effect.
However, since there is no reversal of {\it in} and {\it out} states, this time asymmetry cannot be interpreted as genuine \T violation.

%
%
In this paper we propose and describe a methodology that makes use of the Einstein-Podolsky-Rosen (EPR) entanglement~\cite{ref:EPR} available at 
current B factories to overcome the problem of irreversibility and perform a direct observation of Time Reversal Violation.
The method relies precisely on the possibility for preparing the quantum mechanical individual state of the neutral-\B meson
by the observation of particular decay channels of its orthogonal entangled partner, 
and studying the time evolution of the filtered state of the still living meson.
This strategy allows the interchange {\it in} $\leftrightarrow$ {\it out} states for a given process.
Whereas the basic ideas have been presented 
previously~\cite{ref:bernabeuPLB-NPB} and scrutinized later~\cite{ref:Wolfenstein,ref:QuinnDiscrete,ref:BernabeuDiscrete}, 
here we discuss for the first time the steps to implement these concepts into an actual experiment able to produce the desired results.
In addition, using a 
simulation of data samples of similar size and properties to those currently available at B factories, 
we illustrate the application of the methodology and 
estimate the significance of the expected discovery. The proposed analysis is independent of the underlying theory, \CPT invariant or not, for describing the relevant transitions.

\section{ \boldmath Time Reversal from entangled $\Bz\Bzb$ decays}
\label{sec:entangled}

Bose statistics, for a $\Bz\Bzb$ system with $\C=-$, requires these two mesons to be in an antisymmetric state of the system for any pair of orthogonal individual states. These can be either the $B$ states projected by flavor,
\begin{equation}
\label{eq:initialFlavorstate}
\ket i = \dfrac{1} {\sqrt{2}} [\Bz(t_{1})\Bzb(t_{2})- \Bzb(t_{1})\Bz(t_{2})],
\end{equation}
or the ones projected by \CP final states~\cite{ref:bernabeuJHEP},
\begin{equation}
\label{eq:initialCPstate}
\ket i = \dfrac{1} {\sqrt{2}} [\Bplus(t_{1})\Bminus(t_{2})- \Bminus(t_{1})\Bplus(t_{2})],
\end{equation}
where $t_1$ and $t_2$ are the labels to specify the states ``1'' and ``2'' of each neutral-\B meson by means of the time
of its future decay, with $\dtau = t_2 - t_1 > 0$. 
The antisymmetry remains invariant with the time evolution, including mixing, 
before the first decay at $t_1$.
In Eq.~(\ref{eq:initialCPstate}), \Bminus is the neutral-\B state filtered by its decay to $\jpsi \Kplus$, \Kplus being the neutral-\kaon state filtered by its 
decay to $\pi\pi$, and \Bplus is orthogonal to \Bminus, not connected to $\jpsi \Kplus$. 
Similarly, in Eq.~(\ref{eq:initialFlavorstate}), \Bzb is the neutral-\B state filtered by its decay to $\ellm X$, 
for example 
a semileptonic decay with a negatively charged lepton or 
a hadronic final state 
containing a \Dp or \Dstarp meson.
%
%
We note that treating $\jpsi \KS$, $\KS\to\pi\pi$ final states as $\jpsi \Kplus$, i.e., neglecting \CP violation in the neutral-kaon system,
introduces effects that are small and thus can be neglected. 
Similarly, treating hadronic final states containing $D^{(*)+}$ mesons
as \Bzb, i.e., neglecting \CP violation due to interference from doubly
CKM-suppressed decays, introduces effects small enough that can be considered as corrections.
The same applies to \Bplus as the state connected to $\jpsi \Kminus$ and treated as $\jpsi \KL$,
 and \Bz as the state connected to hadronic final states containing $D^{(*)-}$ mesons.
The identification of \Bminus, \Bplus as those states filtered by the decay to \CP eigenstates is illustrated in Appendix~\ref{sec:basediscussion}.

\begin{figure*}[htb!]
  \begin{center}
    \includegraphics[width=.9\textwidth]{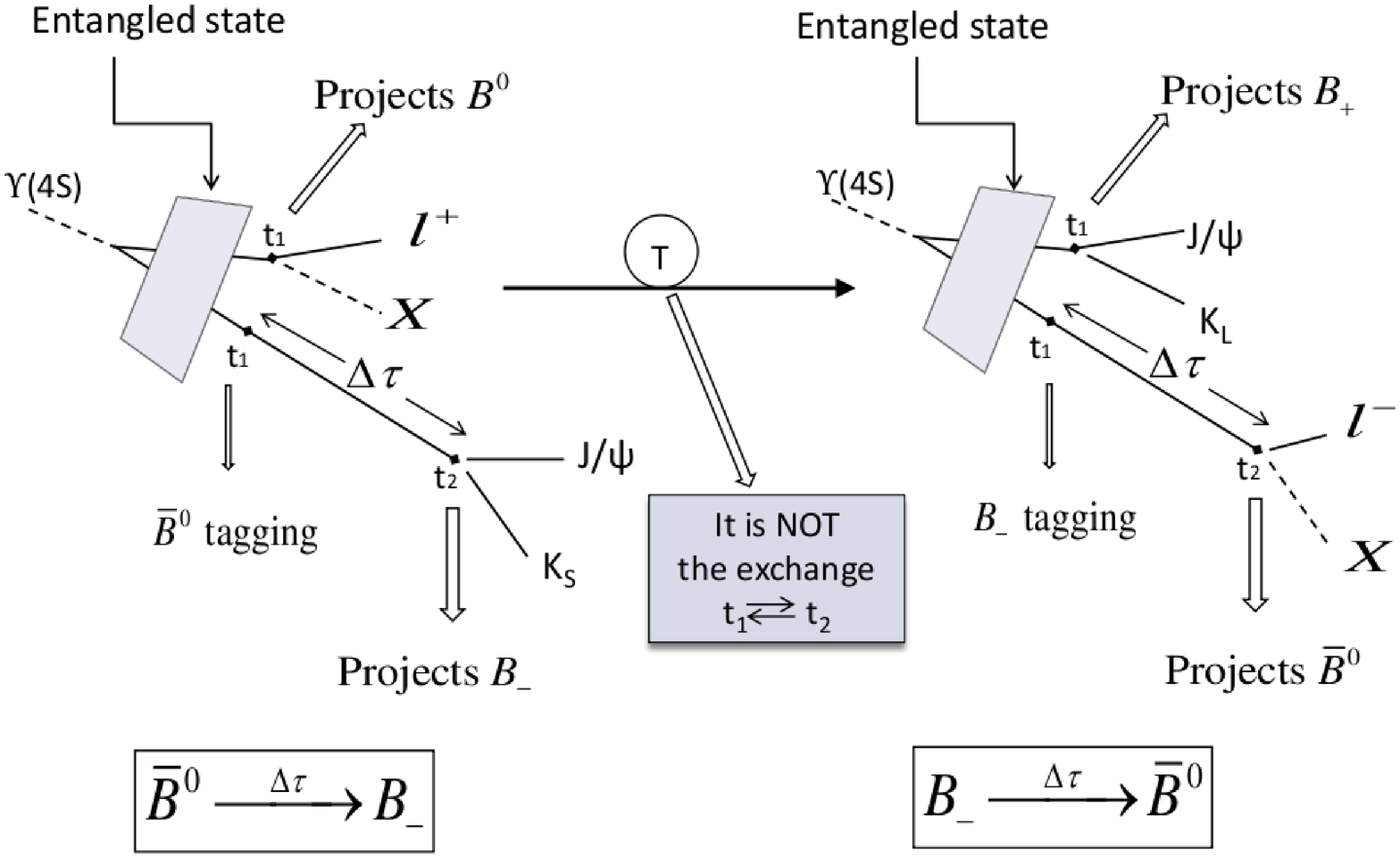}
    \caption{
Sketch of two \T-conjugate transitions in an experimental B factory scheme. 
The observation of the final states associated to the \T-transformed transitions is divided into three well defined steps.
We first observe the decay of one of the entangled \B particles, produced in the $\FourS$ decay,
into a definite flavor (or a definite \CP) decay products at $t_1$, 
preparing the state of the other entangled \B particle, which has not yet decayed at $t_1$, into its orthogonal state.
This tagged \B meson state evolves in time to finally decay at $t_2 > t_1$ into a \CP (or a flavor) final state.
It should be noted that \T asymmetry is clearly different from the \dt ($t_1 \leftrightarrow t_2$) exchange and \CP asymmetries.
In fact, in the former we require to compare the reference transition $\Bzb\to\Bminus$, 
flavor-tagged by $\ellp X$ and decayed to $\jpsi\KS$ ($\ellp X$,$\jpsi\KS$),
to the transition $\Bminus\to\Bzb$, 
\CP-tagged by $\jpsi\KL$ and decayed to $\ellm X$ ($\jpsi\KL$,$\ellm X$),
whereas for \dt and \CP asymmetries the reference decay products must be compared 
to ($\jpsi\KS$,$\ellp X$) and ($\ellm X$,$\jpsi\KS$), respectively.
\label{fig:expprocesses}
    }
  \end{center}
\end{figure*}

Therefore, in addition to the 
{\it flavor tagging} used in standard \CP studies at B factory experiments~\cite{ref:mixingInducedCP-Bs,ref:sin2bPRD}
we can apply a {\it \CP tagging}~\cite{ref:bernabeuJHEP} to one of the \B mesons decaying into 
the \CP-odd final state $\jpsi\KS$, preparing the orthogonal \B meson as \Bplus state
at the initial time $t_1$. Afterwards it
decays at $t_2$ and is reconstructed in the flavor (\Bz or \Bzb) final state of interest ($\ellp X$ or $\ellm X$, respectively).  
This combination of flavor and \CP tags allows to filter initial and final states to compare, for example,
the rate for a \Bzb evolving to \Bminus ($\ellp X$ decay product of the \B partner first, $\jpsi\KS$ final state later)
with the rate for a \Bminus evolving to \Bzb ($\jpsi\KL$ decay product of the \B partner first, $\ellm X$ final state later).
The relation between these \T-conjugated transitions and the reconstructed final states in the experimental B factory scheme
is illustrated in Fig.~\ref{fig:expprocesses}. The comparison between the time evolution of the neutral-\B meson from its preparation as \Bzb until its 
decay as \Bminus, $\Bzb(t_1)\to\Bminus(t_2)$, and its \T transformed $\Bminus(t_1)\to\Bzb(t_2)$ is our proposed test of Time Reversal symmetry.

There are other three independent comparisons between \T-conjugated processes, as summarized in Table~\ref{tab:Ttransformations}. 
A non-vanishing asymmetry in the rates for any pair
of \T-conjugated transitions thus constitutes a direct and independent observation of \T violation, in the sense discussed previously.

\begin{table}[h]
  \begin{center}
    \begin{tabular}{cc|cc}		
      \hline
      \multicolumn{2}{c|}{Reference}  &   \multicolumn{2}{c}{\T-conjugate} \\
      Transition & Final state        &   Transition &  Final state  \\ \hline
      \trule $\Bzb \to \Bminus$ &  $ (\ellp X,\jpsi\KS)$  &  $\Bminus \to \Bzb$ & $(\jpsi\KL,\ellm X)$ \\
      \trule $\Bplus \to \Bz$   &  $ (\jpsi\KS, \ellp X)$ & $\Bz \to \Bplus$    & $(\ellm X,\jpsi\KL)$ \\
      \trule $\Bzb \to \Bplus$  &  $ (\ellp X,\jpsi\KL)$  & $\Bplus \to \Bzb$   & $(\jpsi\KS,\ellm X)$ \\
      \trule $\Bminus \to \Bz$  &  $ (\jpsi\KL,\ellp X)$  &  $\Bz \to \Bminus$  & $(\ellm X,\jpsi\KS)$ \\ \hline
   \end{tabular}
    \caption{Possible comparisons between \T-conjugated transitions and the associated time-ordered decay products in the experimental B factory scheme.
      \label{tab:Ttransformations}}
  \end{center}	
\end{table}

We can also apply this methodology for similar tests of \CP violation and \CPT invariance,
providing a proof that \T non-invariance is compensated by \CP violation.
Tables~\ref{tab:CPtransformations}~and~\ref{tab:CPTtransformations} summarize all the possible
comparisons of \CP- and \CPT-conjugated transitions, together with the corresponding final states. As anticipated, the transitions involved in the experimental tests of \CP and \T symmetries are different. For example, a test of \CP symmetry can be done with the $\jpsi\KS$ final state only. On the contrary, 
a test of \T invariance necessarily involves both $\jpsi\KS$ and $\jpsi\KL$ final states. Furthermore, one may check that none of all comparisons between \T-,  \CP-, or \CPT-conjugated transitions 
in Tables~\ref{tab:Ttransformations},~\ref{tab:CPtransformations}, and~\ref{tab:CPTtransformations}, respectively,
corresponds to exchange of $t_1$ and $t_2$.

\begin{table}[htb!]
  \begin{center}
    \begin{tabular}{cc|cc}		
      \hline
      \multicolumn{2}{c|}{Reference}  &   \multicolumn{2}{c}{\CP-conjugate} \\
      Transition & Final state        &   Transition &  Final state  \\ \hline
      \trule $\Bzb \to \Bminus$ &  $ (\ellp X,\jpsi\KS)$  &  $\Bz \to \Bminus$  & $(\ellm X,\jpsi\KS)$ \\
      \trule $\Bplus \to \Bz$   &  $ (\jpsi\KS, \ellp X)$ &  $\Bplus \to \Bzb$  & $ (\jpsi\KS, \ellm X$) \\
      \trule $\Bzb \to \Bplus$  &  $ (\ellp X,\jpsi\KL)$  &  $\Bz \to \Bplus$   & $(\ellm X,\jpsi\KL)$ \\
      \trule $\Bminus \to \Bz$  &  $ (\jpsi\KL,\ellp X)$  &  $\Bminus \to \Bzb$ &  $ (\jpsi\KL,\ellm X)$ \\ \hline
   \end{tabular}
    \caption{Possible comparisons between \CP-conjugated transitions and the associated time-ordered decay products in the experimental B factory scheme.
      \label{tab:CPtransformations}}
  \end{center}	
\end{table}

\begin{table}[htb!]
  \begin{center}
    \begin{tabular}{cc|cc}		
      \hline
      \multicolumn{2}{c|}{Reference}  &   \multicolumn{2}{c}{\CPT-conjugate} \\
      Transition & Final state        &   Transition &  Final state  \\ \hline
      \trule $\Bzb \to \Bminus$ &  $ (\ellp X,\jpsi\KS)$  &  $\Bminus \to \Bz$  & $(\jpsi\KL,\ellp X)$ \\
      \trule $\Bplus \to \Bz$   &  $ (\jpsi\KS, \ellp X)$ &  $\Bzb \to \Bplus$  & $(\ellp X,\jpsi\KL)$ \\
      \trule $\Bz \to \Bminus$  &  $ (\ellm X,\jpsi\KS)$  &  $\Bminus \to \Bzb$ & $(\jpsi\KL,\ellm X)$ \\
      \trule $\Bplus \to \Bzb$  &  $ (\jpsi\KS,\ellm X)$  &  $\Bz \to \Bplus$   & $(\ellm X,\jpsi\KL)$ \\ \hline
    \end{tabular}
    \caption{Possible comparisons between \CPT-conjugated transitions and the associated time-ordered decay products in the experimental B factory scheme.
      \label{tab:CPTtransformations}}
  \end{center}	
\end{table}

\section{ \boldmath Methodology and T-violating parameters}
\label{sec:method}
We can now proceed to a partition of the complete set of final states with definite flavor and \CP content into 8 pairs,
defined by the first decaying \B at $t_1$ and preparing the tagging state of the still living meson, i.e., \Bz, \Bzb, \Bminus, \Bplus, 
as a function of \mbox{$\dtau=t_2-t_1>0$}. 
Each of these 8 processes has a time-dependent decay rate $g_{\alpha,\beta}^\pm(\dtau)$,
where indexes $\alpha \in \{\ellp,\ellm\}$ and $\beta \in \{\KS,\KL\}$ run over the final states 
with definite flavor ($\ellp X$, $\ellm X$) and \CP eigenstates ($\jpsi\KS$, $\jpsi\KL$), respectively, 
and the upper index $+$ or $-$ indicates 
if the decay to the flavor final state $\alpha$ occurred before or after to the \CP-eigenstate final state $\beta$. 
Thus a $+ \to \ -$ replacement corresponds to \dt exchange, 
which means experimentally the exchange of the two decay products at $t_1$ and $t_2$. 

From only quantum mechanics each decay rate can be written as function of \dtau
%
\begin{eqnarray}
\label{eq:intensity}
g_{\alpha,\beta}^\pm(\dtau)  & \propto  e^{-\Gamma \dtau}\big\{ & C_{\alpha,\beta}^\pm \cos(\dm\dtau)+S_{\alpha,\beta}^\pm \sin(\dm\dtau) + \nonumber \\
   &  &  D_{\alpha,\beta}^\pm \cosh(\Delta \Gamma \dtau)+E_{\alpha,\beta}^\pm \sinh(\Delta \Gamma \dtau)\ \big\},\ \ \ \ 
\end{eqnarray}
%
%
where $\Gamma$ is the average decay width, $\dm$ and $\Delta\Gamma$ are the mass and width differences between the mass eigenstates, 
and $C_{\alpha,\beta}^\pm$, $S_{\alpha,\beta}^\pm$, $D_{\alpha,\beta}^\pm$ and $E_{\alpha,\beta}^\pm$ are generic coefficients. 
This construction makes no assumptions about neither \CPT invariance nor \CP or \T violation.
Assuming $\Delta\Gamma=0$ in the time dependence and renormalizing to the coefficient of the $\cosh(\Delta\Gamma \dtau)$ term, 
Eq.~(\ref{eq:intensity}) simplifies to
\begin{eqnarray}
\label{eq:intensitymod}
g_{\alpha,\beta}^\pm(\dtau) & \propto & e^{-\Gamma \dtau} \big\{ 1+ C_{\alpha,\beta}^\pm \cos(\dm\dtau) + S_{\alpha,\beta}^\pm \sin(\dm\dtau) \big\}.
\end{eqnarray}
The sine term in Eq.~(\ref{eq:intensitymod}) results from the interference between amplitudes with and without mixing,
whereas the cosine term arises from the interference between decay amplitudes with different weak and strong phases.
%
Effects due to any small lifetime difference in the time dependence, 
and the renormalization of all coefficients to the $\cosh(\Delta\Gamma \dtau)$ term 
for each subsample separately, 
introduce small corrections.

It then follows that asymmetries in decay rates for any pair of \T-conjugated transitions (Table~\ref{tab:Ttransformations})
would be apparent through differences between their respective best fit
$S_{\alpha,\beta}^\pm$ and $C_{\alpha,\beta}^\pm$ coefficients.
For example, a significant difference between the \SpLpKs and \SmLmKl coefficients would imply
observation of \T violation.
In general, a net difference of $S_{\alpha,\beta}^\pm$ or $C_{\alpha,\beta}^\pm$ parameters 
between two \T-, \CP-, or \CPT-transformed processes (Tables~\ref{tab:Ttransformations}, ~\ref{tab:CPtransformations}, 
or~\ref{tab:CPTtransformations}) would be a proof of \T, \CP, or \CPT violation, respectively.



The standard \CP violation studies performed by the B factory experiments~\cite{ref:sin2bPRD}
extract a single set of best fit $S$ and $C$ coefficients,
reversing the sign of $S$ under $\dt \leftrightarrow -\dt$,
or $\Bplus \leftrightarrow \Bminus$,
or $\Bz \leftrightarrow \Bzb$ exchanges,
and reversing the sign of $C$ only under $\Bz \leftrightarrow \Bzb$ exchange, 
\begin{eqnarray}
\label{eq:S}
S  & = \dfrac{2 Im(\lambda)}{1+|\lambda|} &=  \SpLpKs = - \SpLmKs = - \SmLpKs =  \SmLmKs = \\ \nonumber
   &   & -\SpLpKl =   \SpLmKl =   \SmLpKl = -\SmLmKl, 
\end{eqnarray}
\begin{eqnarray}
\label{eq:C}
C  & = \dfrac{1-|\lambda|}{1+|\lambda|} &=  \CpLpKs = - \CpLmKs = \CmLpKs = - \CmLmKs = \\ \nonumber
   &   &  \CpLpKl =  -\CpLmKl = \CmLpKl = - \CmLmKl. 
\end{eqnarray}
This construction is valid under the assumptions of \CPT invariance 
 and $\Delta \Gamma =0$~\cite{ref:bernabeuPLB-NPB}. If $\Delta \Gamma =0$, automatically there is no \CP violation in the mixing.
Under these assumptions
the $\dt \leftrightarrow -\dt$ (or equivalently $t_1 \leftrightarrow t_2$) exchange, 
which is not a \T-symmetry operation, becomes related to \T (exchange of the {\it in} and {\it out} neutral-\B states), 
and to \CP (exchange of \Bz and \Bzb states),
$\CP \leftrightarrow \T \leftrightarrow \dt$~\cite{ref:bernabeuNPP}.
In other words, the resulting statement that particle $\leftrightarrow$ antiparticle invariance test is related to the 
exchange $\dt \leftrightarrow -\dt$. In the SM $\lambda=e^{i2\beta}$, therefore 
\begin{eqnarray}
\label{eq:SCinSM}
S  & = & \stwob \\ \nonumber
C  & = & 0.
\end{eqnarray}
As most of the currently available data on neutral-\B mesons are well described by the SM, whether our proposed \T-violating 
observables are also well accounted for in the SM or not, the direct observation of \T violation would provide a proof of principle.


It is now convenient to introduce the asymmetry parameters \DeltaSpmT, \DeltaCpmT 
(and similarly \DeltaSpmCP, \DeltaCpmCP, and \DeltaSpmCPT, and \DeltaCpmCPT),
defined in Table~\ref{tab:new_base},
as the difference of $S_{\alpha,\beta}^\pm$, $C_{\alpha,\beta}^\pm$ coefficients between the corresponding symmetry-transformed processes for two reference \dt-exchanged transitions, for example \SpLpKs and \SmLpKs.
Using these parameters rather than the $S_{\alpha,\beta}^\pm$, $C_{\alpha,\beta}^\pm$ coefficients has the main advantage that the symmetry violation will
be apparent through a non-vanishing value of any of the associated four parameters. In other words,
if $\DeltaSpT \ne 0$ or $\DeltaSmT \ne 0$ or $\DeltaCpT \ne 0$ or $\DeltaCmT \ne 0$, then there is \T violation,
and similarly for \CP and \CPT symmetries. 
Table~\ref{tab:DeltanonT-CP-CPT-violation} summarizes the values and relations among the asymmetry parameters 
in the case of invariance under the three space-time discrete symmetry transformations.
We would like to emphasize that the definition of these parameters requires to choose as reference two \dt-exchanged samples, 
in our case \SpmLpKs since $\jpsi\KS$ events are cleaner and more efficiently reconstructed than $\jpsi\KL$~\cite{ref:sin2bPRD}.

  \begin{table}[htb!]
    \begin{center}
       \begin{tabular}{ c | c | c}		
	 \hline
         Coefficient & Assumed value & Fit value \\ \hline
\trule   \DeltaSpT  = \SmLmKl - \SpLpKs    & $-1.4$ &   $ -1.57\pm 0.15$ \\  
\trule   \DeltaSmT   = \SpLmKl - \SmLpKs   & $\phm1.4$  &   $  \phm1.25\pm 0.19$ \\  
\trule   \DeltaCpT    = \CmLmKl - \CpLpKs  & $\phm0.0$  &   $ -0.07\pm 0.14$ \\
\trule   \DeltaCmT   = \CpLmKl - \CmLpKs   & $\phm0.0$  &   $ -0.09\pm 0.14$ \\ [0.07in] \hline
\trule   \DeltaSpCP = \SpLmKs - \SpLpKs    & $-1.4$ &   $ -1.65\pm 0.11$ \\ 
\trule   \DeltaSmCP  = \SmLmKs - \SmLpKs   & $\phm1.4$  &   $  \phm1.54\pm 0.13$ \\ 
\trule   \DeltaCpCP   = \CpLmKs - \CpLpKs  & $\phm0.0$  &   $  \phm0.03\pm 0.10$ \\
\trule   \DeltaCmCP  = \CmLmKs - \CmLpKs   & $\phm0.0$  &   $ -0.09\pm 0.10$ \\ [0.07in] \hline
\trule   \DeltaSpCPT = \SmLpKl - \SpLpKs   & $\phm0.0$  &   $ -0.25\pm 0.22$ \\  
\trule   \DeltaSmCPT = \SpLpKl - \SmLpKs   & $\phm0.0$  &   $  \phm0.04\pm 0.13$ \\
\trule   \DeltaCpCPT = \CmLpKl - \CpLpKs   & $\phm0.0$  &   $ -0.04\pm 0.15$ \\
\trule   \DeltaCmCPT   = \CpLpKl - \CmLpKs & $\phm0.0$  &   $ -0.06\pm 0.13$ \\ [0.07in] \hline
\trule   \SpLpKs     & $\phm0.7$  &   $  \phm0.92\pm 0.10$ \\
\trule   \SmLpKs     & $-0.7$ &   $ -0.70\pm 0.06$ \\
\trule   \CpLpKs     & $\phm0.0$  &   $  \phm0.08\pm 0.07$ \\ 
\trule   \CmLpKs     & $\phm0.0$  &   $  \phm0.06\pm 0.06$ \\ [0.07in] \hline
       \end{tabular}
       \caption{Definition of the \T-, \CP-, and \CPT-asymmetry parameters. These parameters 
are defined as the differences between the $S_{\alpha,\beta}^\pm$, $C_{\alpha,\beta}^\pm$ coefficients for two reference \dt-exchanged processes and those of the 
corresponding symmetry-transformed transitions. 
In the central column we show the expected values based on the SM \CP violation studies at B factory experiments, 
as given in Eqs.~(\ref{eq:S}) and~(\ref{eq:C}). The right column reports the fit results from one of the 350 simulated experiments
described in Sec.~\ref{sec:simulation}.
\label{tab:new_base}}
    \end{center}	
  \end{table}

\begin{table}[htb!]
\begin{center}
\begin{tabular}{c|c|c} 
\hline
\trule \T invariance            & \CP invariance & \CPT invariance \\ \hline
\trule \DeltaSpT = 0            &\DeltaSpCP = 0            &\DeltaSpCPT = 0             \\
\trule \DeltaSmT = 0            &\DeltaSmCP = 0            &\DeltaSmCPT = 0             \\
\trule \DeltaSpCP = \DeltaSpCPT &\DeltaSpT = \DeltaSpCPT   &\DeltaSpT = \DeltaSpCP    \\
\trule \DeltaSmCP = \DeltaSmCPT &\DeltaSmT = \DeltaSmCPT   &\DeltaSmT = \DeltaSmCP    \\ [0.04in] \hline
\trule \DeltaCpT = 0            &\DeltaCpCP = 0            &\DeltaCpCPT = 0             \\
\trule \DeltaCmT = 0            &\DeltaCmCP = 0            &\DeltaCmCPT = 0             \\
\trule \DeltaCpCP = \DeltaCpCPT &\DeltaCpT = \DeltaCpCPT   &\DeltaCpT = \DeltaCpCP    \\
\trule \DeltaCmCP = \DeltaCmCPT &\DeltaCmT = \DeltaCmCPT   &\DeltaCmT = \DeltaCmCP    \\ [0.04in] \hline
\end{tabular}
\end{center}
\caption{Expected values and relations among the asymmetry parameters under invariance of one of the three discrete space-time symmetry transformations.
}
\label{tab:DeltanonT-CP-CPT-violation}
\end{table}

\section{ \boldmath Simulation study}
\label{sec:simulation}
We show in this section how the proposed methodology can be applied to an actual experiment to produce the desired results.
The experimental samples are generated using a Monte Carlo simulation technique and are similar in size and properties to 
those currently available at B factories and used for standard \CP-violation studies~\cite{ref:sin2bPRD}. 
Thereby this study will provide an up to date estimate of the significance of the expected observation.

\subsection{Generation of samples}
\label{sec:simulation-generation}

Samples of events with a \Bz or \Bzb state (i.e., $\ellp X$ or $\ellm X$) and a \CP-eigenstate 
$\jpsi\KS$, $\psi(2S)\KS$, $\chi_{c1}\KS$, denoted generally as $\c\cbar\KS$,
or $\jpsi\KL$, 
are generated taking into account their relative branching ratios, reconstruction efficiencies and misidentification rates.
We consider about 8000 $\c\cbar\KS$ and 6000 $\jpsi\KL$ events, with purities around 90\% and 60\%, respectively~\cite{ref:sin2bPRD}.
The signal component uses the probability density function (PDF) given in Eq.~(\ref{eq:intensitymod}), 
with values for $\dm=0.507~\ps^{-1}$, $1/\Gamma=1.519~\ps$~\cite{ref:pdg2010}, and the
coefficients $S_{\alpha,\beta}^\pm$ and $C_{\alpha,\beta}^\pm$ as given in Table~\ref{tab:new_base} assuming the SM.
%
The misidentification of \CP final states is accounted for through a signal probability that depends on the two usual kinematic 
variables at B factory experiments, the energy of the $\jpsi\KL$
and the mass of the $\c\cbar\KS$ candidates,
including the different individual background sources~\cite{ref:sin2bPRD}.
%
%
Mistakes in the flavor identification (mistags) 
are included by modifying the time-dependent PDF as follows,
\begin{equation}
h_{\alpha,\beta}^\pm(\dtau) \propto (1-\omega_\alpha)g_{\alpha,\beta}^\pm(\dtau)+\omega_{\alpha}g_{\bar{\alpha},\beta}^\pm(\dtau),
\label{eq:intensitymistag}
\end{equation}
where the $\bar{\alpha}$ index denotes the other flavor state to that given by $\alpha$
and $\omega_\alpha$ represents the fraction of flavor states reconstructed as $\ellm X$ being a \Bz (or as $\ellp X$ being a \Bzb),
with $g_{\alpha,\beta}^\pm(\dtau)$ given by Eq.~(\ref{eq:intensitymod}). 

The experiments have employed flavor identification algorithms that analyze tracks not associated to the completely reconstructed \CP-eigenstate
to assign a given category. These assignments are determined from different inclusive or semi-inclusive signatures, such as isolated primary leptons,
kaons, and pions from \B decays to final states containing $D^*$ mesons, and high momentum charged particles from \B decays.
For this study we use
different categories with 
efficiencies $\epsilon$
(mistags $\omega$) ranging between 
9\% and 17\% (3\% and 42\%), 
with a total effective flavor identification efficiency
of $Q=\epsilon (1-2\omega)^2 \approx 31\%$.

The resolution in the reconstruction of the decay time difference by the experiment introduces a smearing 
such that the observed \dtau might take negative values.
Therefore, the sign of \dt, which indicates if the decay to the flavor final state occurred before or after the \CP final state
(as described in Sec.~\ref{sec:method}),
cannot be used directly to disentangle between $+$ and $-$ events.
To overcome this problem, we describe the observed decay rate as a function of the reconstructed \dt, \dtrec, as
\begin{eqnarray}
{\cal H}_{\alpha,\beta}(\dtrec) & \propto  & h_{\alpha,\beta}^+(\dt)  H(\dt)  \otimes \mathcal{R}(\delta t;\sigma_{\dtrec}) + \nn \\
                            &          & h_{\alpha,\beta}^-(-\dt) H(-\dt) \otimes \mathcal{R}(\delta t;\sigma_{\dtrec}) ,
\label{eq:intensitydt}
\end{eqnarray}
where $H(\dt)$, $H(-\dt)$ are Heaviside step functions, 
$h_{\alpha,\beta}^+(\dt)$, $h_{\alpha,\beta}^-(-\dt)$ are given by Eq.~(\ref{eq:intensitymistag}),
and the symbol $\otimes$ indicates the mathematical convolution with the resolution
function $\mathcal{R}(\delta t;\sigma_{\dtrec})$, where $\delta t = \dtrec - \dt$ and 
$\sigma_{\dtrec}$ is the estimate of the \dtrec uncertainty obtained by the reconstruction algorithms.
%
The first term of the sum is related to $\dt> 0$ (i.e., the tagging \B, identified as $\ellp X$ or $\ellm X$, decayed before the other \B, 
reconstructed as $\ccbar\KS$ or $\jpsi\KL$ state), while the second term is related to $\dt<0$ (the neutral-\B meson decayed to a flavor state later). 
With this construction the distribution ${\cal H}_{\alpha,\beta}(\dtrec)$ for $\dtrec>0$ will be dominated by 
$\dt>0$ events, but will also contain events with $\dt<0$ due to the limited \dt resolution. Similarly, 
the distribution ${\cal H}_{\alpha,\beta}(\dtrec)$ for $\dtrec<0$ 
will contain predominantly events having $\dt<0$ with contribution from $\dt>0$ events.
The need of distinguishing between $\dt<0$ and $\dt>0$ in the presence of \dt resolution is one of the main
complications of this study in comparison to the standard \CP violation analyses at B factories. 
This is the reason of the Heaviside step function in Eq.~(\ref{eq:intensitydt})
and the $\pm$ index in the upper part of the $C_{\alpha,\beta}^\pm$ and $S_{\alpha,\beta}^\pm$ parameters.

The resolution function $\mathcal{R}(\delta t;\sigma_{\dtrec})$
is modeled by the sum of three Gaussian functions~\cite{ref:sin2bPRD} 
with means and widths proportional to $\sigma_{\dtrec}$. 
The mean scale factor for the main component (which includes about 90\% of the events)
ranges between $0$ for flavor identification categories containing isolated primary leptons to
$-0.2$ for those categories without leptons, to account for the typical
offset in the resolution function observed by the experiments. 

\subsection{Sensitivity results}
\label{sec:simulation-sensitivity}

We generate a total of 350 experiments containing all 8 pairs of flavor-\CP events, each of which is then fitted using 
an unbinned maximum likelihood procedure with the same PDF as that employed for the generation of the samples. 
Each fit determines the best 8 asymmetry parameters of Table~\ref{tab:new_base},
with all other parameters of the PDF, like those describing the resolution function, mistags and backgrounds, kept fixed.
This is a realistic scenario for an experimental analysis since all these parameters are usually determined
by the experiments using either control samples or detailed simulations of the detectors.
In order to increase sensitivity and account for the significantly different reconstruction efficiencies between
the different \CP final states (especially between $\jpsi\KL$ and other events) the normalization of the 
PDF in the fit is performed separately for the different \CP final states, but 
simultaneously for $\ellp X$, $\ellm X$, $\dtrec>0$ and $\dtrec<0$ events.

Table~\ref{tab:toymcgoldresults-signal+background} summarizes the standard deviation (root mean square, or r.m.s)
of the residual distribution (fitted minus generated values), the mean of the fit uncertainty, and the
standard deviation of the fit uncertainty, for each of the asymmetry parameters from the 350 experiments.
The expected sensitivities for the \T-violating parameters are $0.15$, $0.20$, $0.14$ and $0.15$,
for \DeltaSpT, \DeltaSmT, \DeltaCpT, and \DeltaCmT, respectively. While the sensitivity for \DeltaCpT and \DeltaCmT is similar, 
it is significantly worse for \DeltaSmT in comparison to \DeltaSpT, since the associated decay rates involve time-dependent 
distributions with larger tails. 
The uncertainties 
obtained directly by the fit (parabolic errors) provide a good estimator of the true resolution
given by the r.m.s of the residual distributions, while no statistically significant biases 
are observed,
concluding that all asymmetry parameters have good Gaussian behavior.
To illustrate the case of an actual experimental analysis, in Table~\ref{tab:new_base} we give the complete fit results from one of the
simulation experiments. As it can be observed, all 
asymmetry parameters are consistent with the generated values (indicated in the same Table), with fit errors
in good agreement with the expected sensitivities.



\begin{table}[!htbp]
  \begin{center}
    \begin{tabular}{l|ccc}
      \hline
Parameter               & r.m.s residual          & mean error      & r.m.s error             \\
\hline                                                                                  
\trule $\DeltaSpT$             & 0.149                 & 0.148                 & 0.009                 \\ 
\trule $\DeltaSmT$             & 0.201	                & 0.201                 & 0.009                 \\
\trule $\DeltaCpT$             & 0.139	                & 0.134                 & 0.008                 \\
\trule $\DeltaCmT$             & 0.150	                & 0.141                 & 0.009                 \\  [0.04in] \hline
\trule $\DeltaSpCP$            & 0.120	                & 0.106                 & 0.005                 \\  
\trule $\DeltaSmCP$            & 0.123	                & 0.120                 & 0.005                 \\  
\trule $\DeltaCpCP$            & 0.103	                & 0.094                 & 0.004                 \\
\trule $\DeltaCmCP$            & 0.103	                & 0.098                 & 0.003                 \\ [0.04in] \hline
\trule $\DeltaSpCPT$           & 0.209	                & 0.213                 & 0.017                 \\
\trule $\DeltaSmCPT$           & 0.137	                & 0.131                 & 0.009                 \\
\trule $\DeltaCpCPT$           & 0.143	                & 0.147                 & 0.009                 \\ 
\trule $\DeltaCmCPT$           & 0.128	                & 0.128                 & 0.007                 \\ [0.04in] \hline
\trule $\SpBzKs$               & 0.101	                & 0.090                 & 0.004                 \\
\trule $\SmBzKs$               & 0.060	                & 0.057                 & 0.003                 \\
\trule $\CpBzKs$               & 0.069	                & 0.066                 & 0.002                 \\
\trule $\CmBzKs$               & 0.058	                & 0.057                 & 0.002                 \\ [0.07in]
\hline
    \end{tabular}
    \caption{Root mean square (r.m.s.) of the residual distribution (fit minus generated values), 
mean of the fit uncertainty, and the r.m.s of the fit uncertainty, for each of the asymmetry parameters from the 350 simulation experiments.
\label{tab:toymcgoldresults-signal+background}}
  \end{center}
\end{table}

\subsection{Asymmetries}

The difference in the rates for any pair of symmetry-transformed transitions normalized to their sum is usually used as
observable to probe the symmetry violation. In practice, we construct raw asymmetries using 
the number of events for each pair of transitions in bins of $\dtplus \equiv |\dtrec|$, normalized to the total 
number of events of the given subsample. This normalization is particularly relevant for \T (and \CPT) asymmetries 
since it involves comparison of $\c\cbar\KS$ and $\jpsi\KL$ states, which overall have different reconstruction 
efficiencies.
Since for a given discrete symmetry there are four possible comparisons between conjugated processes, we have four independent asymmetries. 
%
%
For example, for the $\Bzb \to \Bminus$ reference transition the raw \T asymmetry would explicitly be defined as
\begin{equation}
A_{\T}(\dtplus) = \dfrac{ {\cal H}_{\ellm,\KL}^-(\dtplus) - {\cal H}_{\ellp,\KS}^+(\dtplus) }
                       { {\cal H}_{\ellm,\KL}^-(\dtplus) + {\cal H}_{\ellp,\KS}^+(\dtplus) }~,
\end{equation}
where ${\cal H}_{\alpha,\beta}^\pm(\dtplus) = {\cal H}_{\alpha,\beta}(\pm\dtrec) H(\dtrec)$.
Neglecting mistag 
and proper-time resolution effects, and 
assuming Eqs.~(\ref{eq:S}) and~(\ref{eq:C}) in the denominator for simplicity, the raw \T-asymmetry becomes proportional to $(\DeltaCpT,\DeltaSpT)$,
\begin{equation}
A_{\T}(\dtplus) \approx \frac{\DeltaCpT}{2} \cos\dm\dtplus + \frac{\DeltaSpT}{2} \sin\dm\dtplus.
\end{equation}
The other three \T asymmetries are constructed similarly using the association between transitions and time-ordered decay products
for reference and \T-conjugate processes given in Table~\ref{tab:Ttransformations},
and are proportional to 
$(\DeltaCmT,\DeltaSmT)$,
$(\DeltaCmCP-\DeltaCmCPT,\DeltaSmCP-\DeltaSmCPT)$,
and
$(\DeltaCpCP-\DeltaCpCPT,\DeltaSpCP-\DeltaSpCPT)$,
respectively.
The raw \CP and \CPT asymmetries are constructed 
following a similar procedure.

It should be noted, that with the methodology proposed, these asymmetries are only used with the purpose of
illustrating the symmetry violation effect, through direct comparisons of the time-dependent raw asymmetries from data and the
projections of the best fit results to the decay rates
when we allow for both non-invariance and invariance under the symmetry transformation. 



\begin{figure}[htbp]
\begin{center}
  \includegraphics[width=0.42\textwidth]{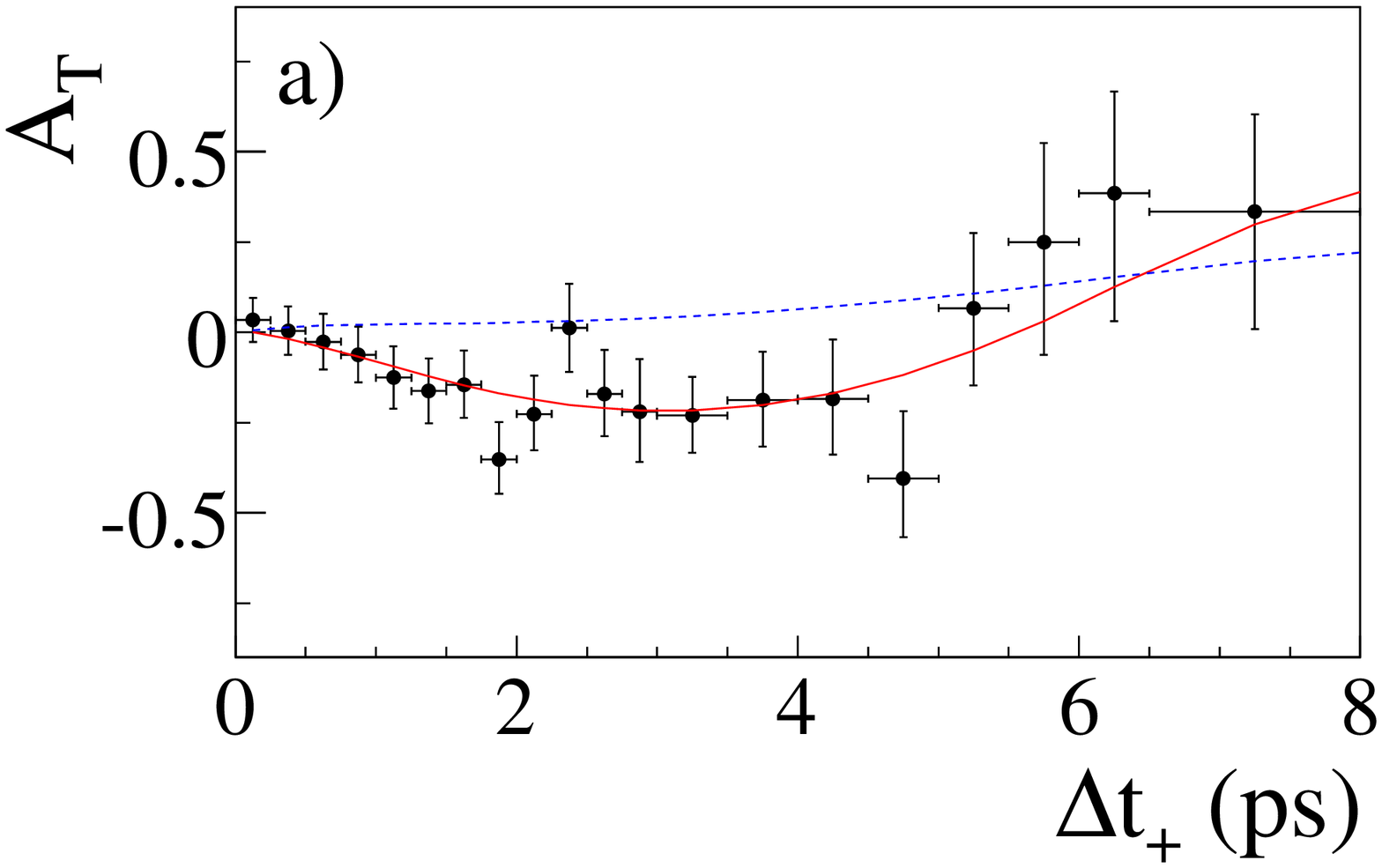} 
  \includegraphics[width=0.42\textwidth]{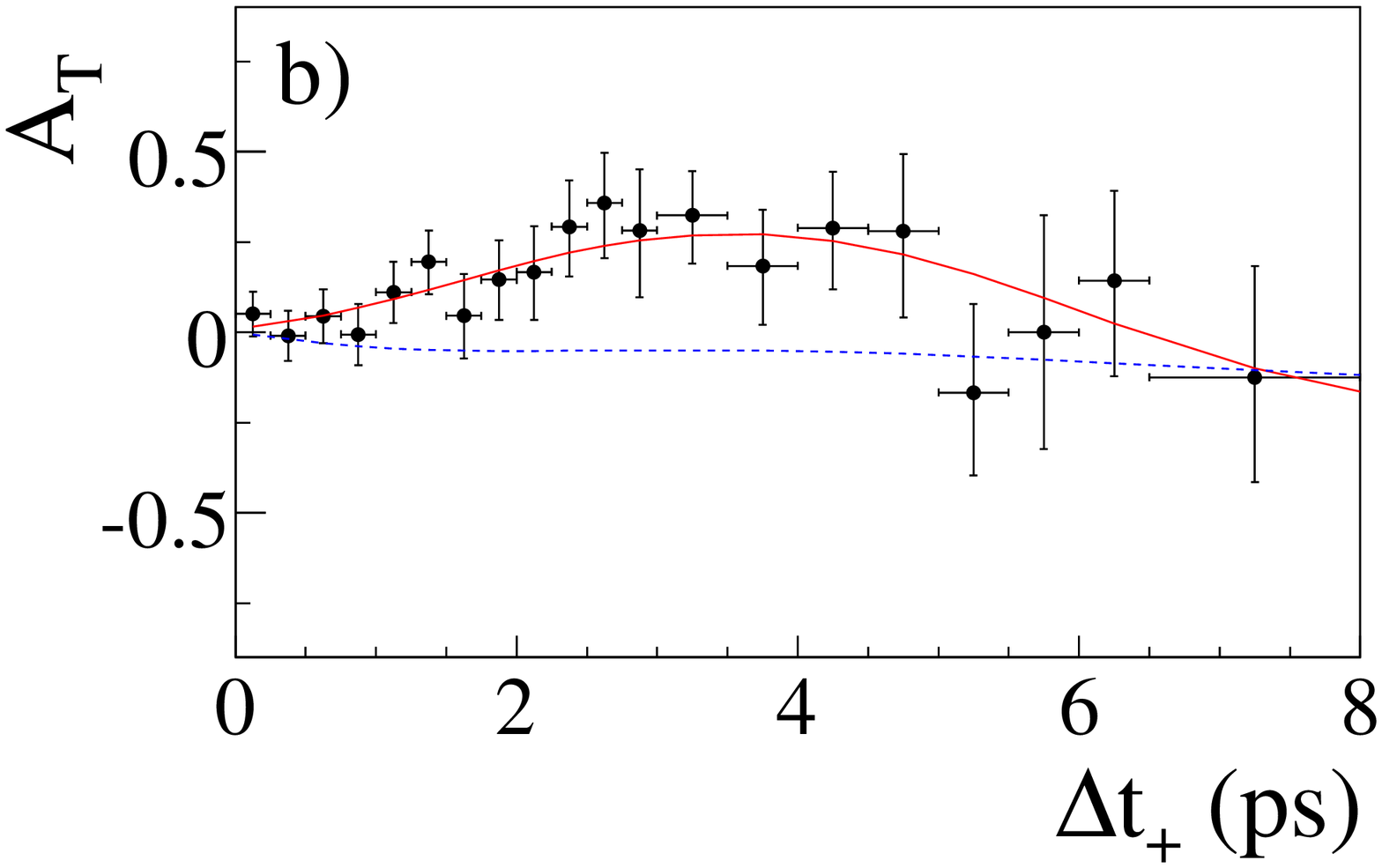} 
  \includegraphics[width=0.42\textwidth]{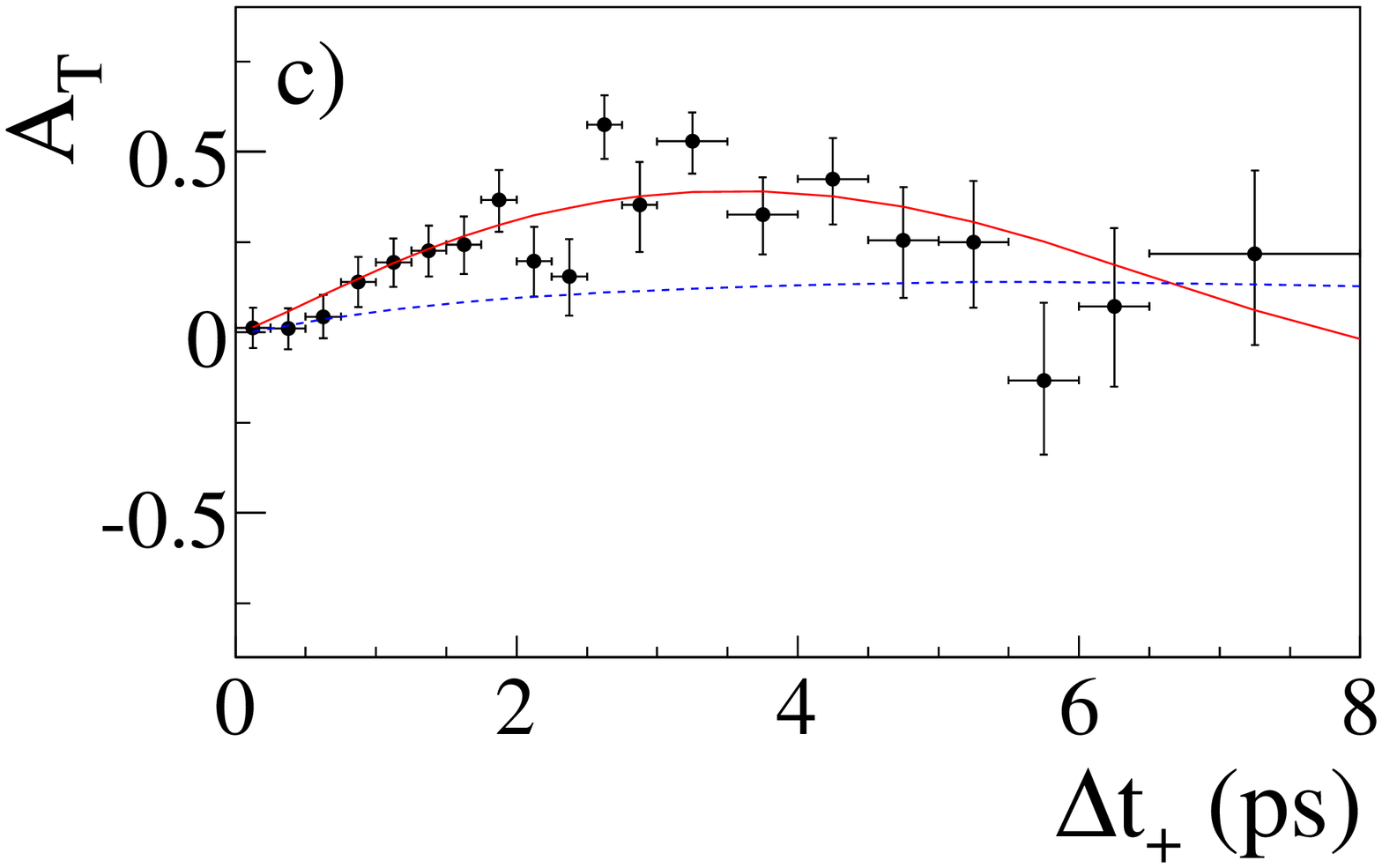}
  \includegraphics[width=0.42\textwidth]{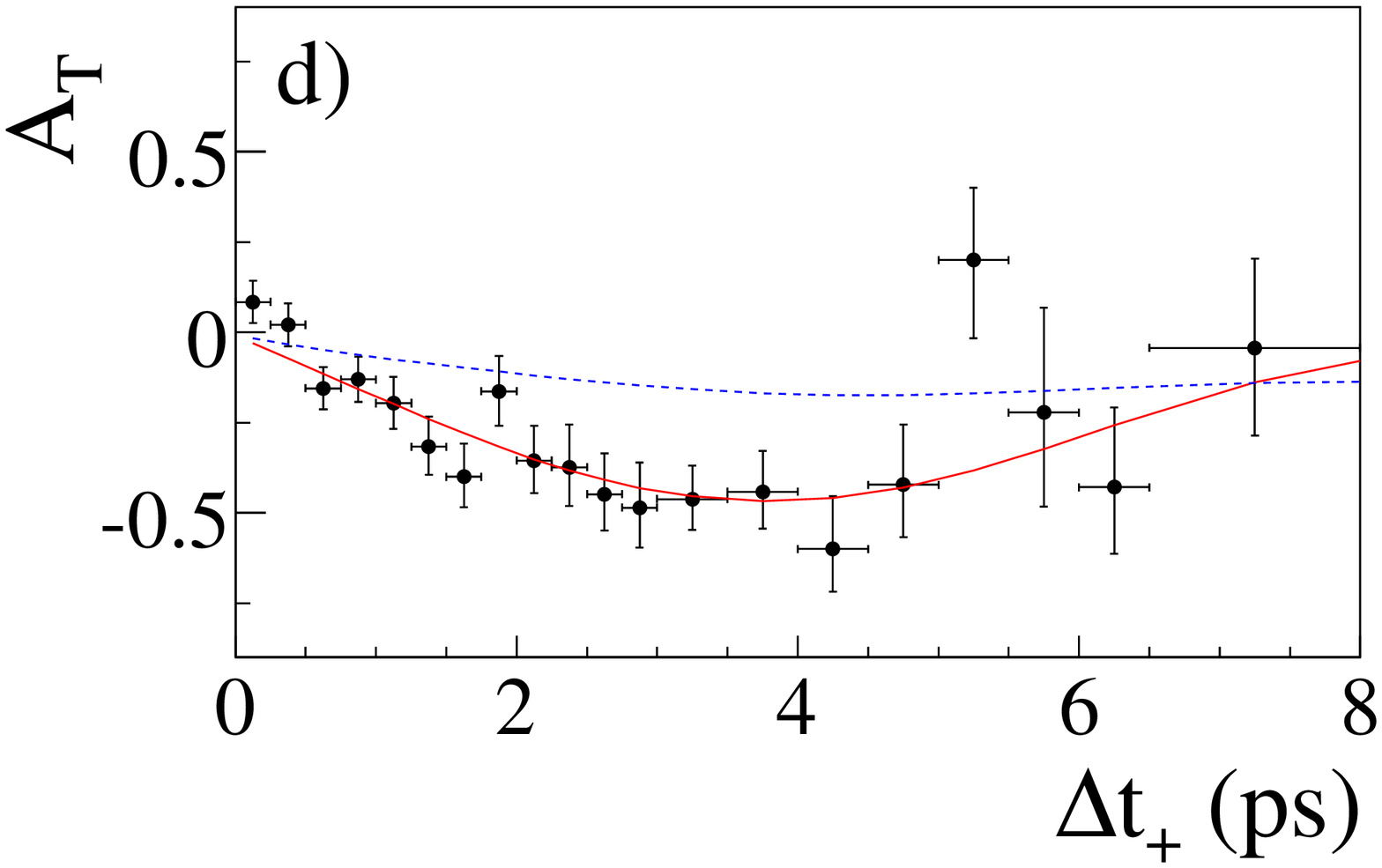} 
\caption{The four independent raw \T-asymmetries corresponding to the four possible comparisons between \T-conjugated and reference
transitions
a) $\Bzb \to \Bminus$ $(\ellp X,\c\cbar\KS)$,
b) $\Bplus \to \Bz$ $(\c\cbar\KS, \ellp X)$,
c) $\Bzb \to \Bplus$ $(\ellp X,\jpsi\KL)$,
and d) $\Bminus \to \Bz$ $(\jpsi\KL,\ellp X)$ of Table~\ref{tab:Ttransformations}, for one of the 350 simulation experiments and 
combining flavor categories with low mistag (isolated leptons and kaons), 
for a signal enriched region. The points with error bars represent the simulated data, 
the solid (red) curves represent the projections of the best fit results,
while the dashed (blue) curves represent the projection of the best fit assuming \T invariance.
%
%
}
\label{fig:sig+bkgtoymcfit-Tasym}
\end{center}
\end{figure}

Figure~\ref{fig:sig+bkgtoymcfit-Tasym} shows the four independent raw \T-asymmetries from data for one of the 350 simulation experiments,
overlaid with the projection of the best fit results together with the projection of the best fit under the assumption of \T invariance.
The \T invariance hypothesis implies to impose to our most general model in Eq.~(\ref{eq:intensitymod}) used in the former case the 8 
restrictions given in the left column of Table~\ref{tab:DeltanonT-CP-CPT-violation}.
This reduces the number of signal parameters left free in the fit from 16 to 8. 
The non-flat shape of the asymmetry curve with \T invariance imposed is due to the offset of the resolution function
\mbox{discussed in Sec.~\ref{sec:simulation-generation}}.

\subsection{Significance of results}

The significance of \T non-invariance can be evaluated by the variation of likelihood 
($s^2_{\T}$)
in the space of asymmetry parameters. This variation is determined by the difference between the negative log-likelihood 
of the best fit with the most general model ($\ln{\cal L}$)
and that of the best fit imposing the 8 restrictions related to \T invariance listed in 
Table~\ref{tab:DeltanonT-CP-CPT-violation} (\T-invariance point, $\ln{\cal L}_{\T}$), 
\bea
s^2_{\T} & \equiv & \twoDLL_{\T} = -2 \left( \ln{\cal L}_{\T} - \ln{\cal L} \right).
\eea
In the limit that the likelihood function takes a Gaussian shape in the space of asymmetry parameters,
which is a good approximation in our case as discussed in Sec.~\ref{sec:simulation-sensitivity} and 
summarized in Table~\ref{tab:toymcgoldresults-signal+background}, 
$s^2_{\T}$ is properly described by a $\chi^2$ distribution with $\nu=8$ degrees of freedom. 
The confidence level (\CL) can then be obtained by computing the cumulative $\chi^2$ probability that the value 
$s^2_{\T}$ is exceeded~\cite{ref:pdg2010}.

For the same simulation experiment shown in Figure~\ref{fig:sig+bkgtoymcfit-Tasym} the variation $s^2_{\T}$ is 353 units,
which corresponds to a \CL equivalent to about $18$ standard deviations (here we have adopted the convention 
that $1,2,3,...$ standard deviations in 8 dimensions have $1-\CL=0.3173,4.55\times10^{-2},2.7\times10^{-3}$,...). 
This is the figure of merit that establishes the degree of inconsistency between the best fit result
and the \T invariance point, i.e., the significance 
that any of the 8 conditions in the left column of Table~\ref{tab:DeltanonT-CP-CPT-violation} is not satisfied,
or equivalently, 
how significant is the combined difference between the solid (red) and dashed (blue) curves for the four 
asymmetries in Figure~\ref{fig:sig+bkgtoymcfit-Tasym}.

The significance of \CP and \CPT non-invariance can be evaluated similarly.
For the same experiment as previously, the variations in $s^2$ for the \CP- and \CPT-invariance points are found 536 and 2 units,
corresponding to about $22$ and less than $1$ standard deviations, respectively. These values are consistent with 
\T non-invariance compensating \CP violation so that \CPT remains invariant, in agreement with the input parameters used in the generation 
of the simulated experiment.



\section{ \boldmath Conclusions}
\label{sec:conclusions}

This work concerns the study of microscopic Time Reversal Violation in the fundamental laws of physics.
The observed time asymmetries in current macroscopic and microscopic phenomena are not connected to this problem
and they can occur in theories with exact \mbox{\T invariance}.

A direct evidence for \T violation means an experiment that, considered by itself, clearly demonstrates \T violation independent of the 
results for \CP violation. For transition processes, the antiunitarity of the operator implementing the symmetry 
transformation implies that a genuine test of \T violation needs an asymmetry under the interchange of {\it in} and {\it out} states in 
the dynamical evolution of the system. 
Nowadays, there is no experimental result providing direct evidence of genuine \T violation.
The measured asymmetry among the probabilities $\Kz\to\Kzb$ and $\Kzb\to\Kz$ 
cannot be interpreted as such since, being \CPT-even transitions, 
\CP and \T are experimentally identical, no matter whether there is \CPT invariance or not. 
Other experiments that could provide evidence involve a non-vanishing expected value of
\T-odd observables for stationary, non-degenerate states, like electric dipole moments,
not yet observed. 
For unstable particles, where we might expect large effects, the associated irreversibility seems to prevent a direct test of \T symmetry 
through the exchange of {\it in} and {\it out} states.

This paper discusses how to overcome this irreversibility problem and defines the precise steps for implementing an actual experiment able to 
obtain direct evidence of genuine \T violation in the time evolution of a neutral-\B meson, taking place
between the two decays of the $\Bz\Bzb$ system produced at B factories.
The essential ingredients are the quantum mechanical entanglement imposed by the EPR correlation,
in particular, the first decay of one \B prepares the quantum state of its living partner,
and the experimental ability to unfold the ordering and difference of the two decay times.
Identifying the \B decays into definite flavor ($\ellp X$ or $\ellm X$) and \CP-eigenstates ($\jpsi\KS$ or $\jpsi\KL$), 
the difference between the decay rates for ($\ellp X$, $\jpsi\KS$) and ($\jpsi\KL$, $\ellm X$) transitions at the 
same decay time difference is our signature of \T violation.
There are three other independent signatures which can be constructed from the
eight time-ordered pairs of \B decays. 
These proofs are independent of the results for \CP violation, as shown by the fact that, given a reference
($\ellp X$, $\jpsi\KS$) for example, the \T-, \CP-, \CPT-transformed, and \dt-exchanged processes are all
experimentally different.

To demonstrate the feasibility of the proposed methodology and how it can be applied for a real data experimental analysis,
we have performed a realistic Monte Carlo simulation of data samples with similar size and properties to those currently available at B 
factories, taking into account 
misidentification of \CP eigenstates, mistakes in the flavor identification and resolution in the 
reconstruction of the decay time difference. 
The need of resolving simultaneously the ordering and difference of the two decay times in the presence of 
resolution is one of the main challenges in comparison to standard \CP violation studies.
For the purpose of illustrating the \T-violating effect, we have presented a direct comparison of 
the time-dependent raw asymmetries from data and the projections of the best fit results to the decay rates 
with and without \T non-invariance.

Using a large number of generated experiments, we have inferred the expected sensitivities
of the asymmetries in the parameters which determine the time-dependent decay rates,
together with the expected significance of \T non-invariance.
A large significance, equivalent to about 18 standard deviations, is foreseen with the currently available data,
providing the expectation of a discovery of direct, genuine Time Reversal Violation in the time evolution of a neutral-\B meson.

\section{ \boldmath Acknowledgments}
\label{sec:acknowl}
We would like to thank the discussions maintained with several colleagues, 
especially Helen R. Quinn, Lincoln Wolfenstein, Klaus Schubert, Antonio di Domenico, Roland Waldi and
Chih-hsiang Cheng.
This work is supported by the Spanish and Generalitat Valenciana grants
FPA-2008-02878. FPA-2008-03917, PROMETEO-2008/004, and PROMETEO-2010/056.


\appendix
\section{The \ket\Bminus\ and \ket\Bplus\ states}
\label{sec:basediscussion}
The $\ket\Bminus$, $\ket\Bplus$ states are experimentally identified as those filtered by the observation of the decay to definite \CP eigenstates. The only requirement needed for the analysis is their existence, independent of the underlying theory (\CPT invariant or not), with the ortogonality property $\langle\Bminus|\Bplus\rangle=0$. In this Appendix we construct these states explicitely.

Neutral kaons decaying inside the geometrical acceptance of detectors surrounding their production point are 
usually reconstructed through their decay to two pions. Thus we can first define $\ket\Bminus$ as the state filtered by the decay into $\jpsi\Kplus$, \Kplus being the neutral-\kaon state decaying to $\pi\pi$, a pure \CP-odd state. 
The state $\ket\Bplustilde$ is then defined as
the state orthogonal to $\ket\Bminus$,
$\langle\Bplustilde|\Bminus\rangle=0$, which cannot decay into the $\jpsi\pi\pi$ final state,
$\langle\jpsi\pi\pi|T|\Bplustilde\rangle=0$, where $T$ is the transition operator. 
The state $\ket\Bplustilde$ can be written in terms of flavor eigenstates as
\begin{equation}
|\Bplustilde\rangle \equiv {\rm\widetilde N_+} \left[| \Bz\rangle - \alpha|\Bzb \rangle \right],
\end{equation}
where $\alpha=\dfrac{\langle \jpsi\pi\pi |T |\Bz\rangle} {\langle \jpsi\pi\pi |T |\Bzb\rangle}$
and ${\rm\widetilde {N}_+}$ is a normalization constant. Since $\ket\Bminus$ is orthogonal 
to $\ket\Bplustilde$ it then follows
\begin{equation}
|\Bminus\rangle = {\rm N_-} \left[| \Bzb\rangle + \alpha^* |\Bz \rangle \right].
\end{equation}

Analogously, we define $\ket\Bplus$ as the state filtered by the decay into $\jpsi\KL$, a \CP-even state up to ${\cal O}(10^{-3})$ due to
\CP violation in the neutral-kaon system. 
We note that we could have defined $\ket\Bplus$ through its decay to $\jpsi\Kminus$, $\Kminus \to \pi^0\pi^0\pi^0$, in this case a pure \CP-even state,
but this final state cannot be reconstructed since long-lived neutral kaons tend to interact hadronically inside the detectors
before they can undergo decay. Of course, both definitions become operationally identical when \CP violation in neutral-kaons is neglected.
The state $\ket\Bminustilde$, defined as its orthogonal state,
$\langle\Bminustilde|\Bplus\rangle=0$ and $\langle\jpsi\KL|T|\Bminustilde\rangle=0$, is
\begin{equation}
|\Bminustilde\rangle \equiv {\rm\widetilde N_-} \left[| \Bz\rangle - \beta |\Bzb \rangle \right],
\end{equation}
where $\beta=\dfrac{\langle \jpsi\KL |T |\Bz\rangle} {\langle \jpsi\KL |T |\Bzb\rangle}$.
Therefore the state $\ket\Bplus$ can be explicitly written as
\begin{equation}
|\Bplus\rangle = {\rm N_+} \left[| \Bzb\rangle + \beta^* |\Bz \rangle \right].
\end{equation}


Let us note that here we keep separate the definitions of 
the states $\ket\Bminus$ and $\ket\Bplus$,
which are observed through their decays to the $\jpsi\pi\pi$ and $\jpsi\KL$ final states,
from the states $\ket\Bminustilde$ and $\ket\Bplustilde$, 
produced exploiting the EPR correlations in the entangled \BB system.
The two bases are the same when the orthogonality condition $\langle\Bminus|\Bplus\rangle=0$ is fulfilled.
As $\alpha$ and $\beta$ correspond to opposite \CP final states, we have $\alpha\beta^*=-1$
when only one weak amplitude is responsible of the decay ($|\alpha|=|\beta|=1$). 
Consequently as all these considerations apply to our experimental framework, 
\begin{equation}
\label{eq:orth}\langle\Bminus|\Bplus\rangle = 1+\alpha\beta^* = 0~,
\end{equation} 
which was to be demonstrated.
%
%

\end{document}